\documentclass[wrr]{agu2001}

\usepackage{multirow}
\usepackage{amsmath}
\usepackage{amssymb}
\usepackage{graphicx}

\authorrunninghead{RIBATET ET AL.}

\titlerunninghead{Reversible Jump Techniques in Regional Flood
  Frequency Analysis}

\authoraddr{M. Ribatet,
  Unit\'e de Recherche HH, Cemagref Groupement de Lyon, 3bis quai
  Chauveau CP220, 69336 Lyon Cedex 09,
  FRANCE. (ribatet@lyon.cemagref.fr)}

\paperid{2006WR005525}
\issuenumber{8}
\articleid{43}{}
\journalid{2007}{}

\begin{document}
\title{Usefulness of the Reversible Jump Markov Chain Monte Carlo
  Model in Regional Flood Frequency Analysis}

\author{M. Ribatet, \altaffilmark{1,2}
  E. Sauquet, \altaffilmark{1}
  JM. Gr\'esillon, \altaffilmark{1}
  and T.B.M.J. Ouarda \altaffilmark{2}}

\altaffiltext{1}
{CEMAGREF Lyon, Unit\'e de Recherche Hydrologie-Hydraulique, 3 bis
  quai Chauveau, CP220, 69336 Lyon cedex 09, FRANCE}

\altaffiltext{2}
{INRS-ETE, University of Qu\'ebec, 490, de la Couronne Qu\'ebec, Qc,
  G1K 9A9, CANADA.} 

\begin{abstract}
  Regional flood frequency analysis is a convenient way to reduce
  estimation uncertainty when few data are available at the gauging
  site. In this work, a model that allows a non null probability to a
  regional fixed shape parameter is presented. This methodology is
  integrated within a Bayesian framework and uses reversible jump
  techniques. The performance on stochastic data of this new estimator
  is compared to two other models: a conventional Bayesian analysis
  and the index flood approach. Results show that the proposed
  estimator is absolutely suited to regional estimation when only a
  few data is available at the target site. Moreover, unlike the index
  flood estimator, target site index flood error estimation seems to
  have less impact on Bayesian estimators. Some suggestions about
  configurations of the pooling groups are also presented to increase
  the performance of each estimator.
  \vspace{10pt}
\end{abstract}


\begin{article}
  
  \section{Introduction}
  \label{sec:intro}

  Extreme value theory is now widely applied when modeling block
  maxima or exceedences over a threshold is of interest. In
  particular, the Generalized Pareto Distribution (\textbf{GPD})
  describes the limiting distribution of normalized excesses of a
  threshold as the threshold approaches the endpoint of the variable
  \citep{Pickands1975}. The GPD has a distribution function defined
  by:

  \begin{equation}
    \label{eq:gpd}
    G(x; \mu, \sigma, \xi) = 1 - \left[1 + \frac{\xi(x - \mu)}{\sigma}
    \right]^{-1/\xi}, x > \mu, 1 + \frac{\xi(x -
      \mu)}{\sigma} > 0
  \end{equation}
  where $\sigma>0$, $\xi \in \mathbb{R}$. $\mu, \sigma$ and $\xi$ are
  respectively the location, scale and shape parameters.

  Thus, when extreme values must be estimated, this approximation is
  frequently used. Most applications based on this result are related
  to environmental sciences, as extreme wind speed \citep{Payer2004},
  extreme sea level \citep{Bortot2000,Pandey2004} or extreme river
  discharge \citep{Northrop2004}.

  However, one must often deal with small samples and large
  uncertainties on estimation. Several publications point out the
  problem of the shape parameter estimation. This parameter is of
  great interest as it determines the tail behaviour of the
  distribution. Therefore, many authors analyzed the performance of
  particular estimators given a specified range for the shape
  parameter: \cite{Rosbjerg1992} for the method of moments;
  \cite{Coles1999b} for the maximum likelihood; \cite{Hosking1987} for
  the probability weighted moments; \cite{Juarez2004} for the minimum
  density power divergence estimator; \cite{Martins2000} for a
  proposed generalized maximum likelihood. However, these results
  provide the most accurate estimator given the shape parameter; which
  is never the case in practice. Therefore, \cite{Park2005} introduced
  a systematic way of selecting hyper-parameters for his proposed
  generalized maximum likelihood estimator.

  All these approaches only deal with information from the target site
  sample. However, it is frequent in hydrology to perform a Regional
  Frequency Analysis (\textbf{RFA}). Traditional RFA consists of two
  steps: (a) delineation of homogeneous regions i.e. a pooling group
  of stations with similar behaviour; (b) regional estimation i.e.
  estimate target site distribution from the regional information.

  More recently, Bayesian approaches have been applied with success to
  incorporate regional information in frequency analysis
  \citep{Coles1996,Northrop2004,Seidou2006,Ribatet2007a}. Empirical
  Bayesian estimators have also been proposed
  \citep{Kuczera1982,Madsen1997c}.  One of the advantages of these
  approaches is to distinguish the at site information from the other
  sites data in the estimation procedure.  This is an important point
  as, no matter how high the homogeneity level may be, the only data
  which represents perfectly the target site is obviously the target
  site one. Thus, the whole information available is used more
  efficiently. In addition, according to \citet{Ribatet2007a}, the
  Bayesian approaches allow to relax the scale invariance property
  required by the most applied RFA model, that is, the index flood
  \citep{Dalrymple1960}.

  However, a preliminary study on simulated data showed that the
  approach developed by \citet{Ribatet2007a} may lead to unreliable
  estimates for larger return periods ($T > 20$ years) when small
  samples are involved. This poor performance is mainly due to the
  large variance on the shape parameter estimation. Consequently, for
  such cases, attention must be paid to the regional estimation
  procedure for the shape parameter.

  The basis of our new development was formerly proposed by
  \citet{Stephenson2004a}. They use reversible jump Markov chain Monte
  Carlo techniques \citep{Green1995} to attribute a non null
  probability to the Gumbel case. Therefore, realizations are not
  supposed to be Gumbel distributed, but have a non null probability
  to be Gumbel distributed. An application to extreme rainfall and
  sea-level is given. In this work, this approach is extended to take
  into account a regional shape parameter, not only the
  Gumbel/Exponential case, within a RFA framework. The reversible jump
  technique allows to focus on a ``likely'' shape parameter value
  given by the hydrological relevance of the homogeneous region.
  Thus, this approach may reduce the shape parameter variance
  estimation while relaxing the scale invariance property.

  The main objectives of this article is first to present new
  developments in the methodology proposed by \citet{Stephenson2004a}
  required for a RFA context; second to assess the quality of two
  Bayesian models based on the index flood hypothesis: the regional
  Bayesian model proposed by \cite{Ribatet2007a} ($\mathbf{BAY}$) and
  the new proposed Bayesian approach applying reversible jumps Markow
  chains ($\mathbf{REV}$).  They are compared to the classical index
  flood approach of \cite{Dalrymple1960} ($\mathbf{IFL}$). The
  assessment is developed through a stochastic generation of regional
  data performed in order to obtain realistic features of homogeneous
  regions. Detailing the index flood concept is out of the scope of
  this article. Estimation procedure can be found in
  \cite{Hosking1997}.

  The paper is organized as follows. The next two sections concentrate
  on methodological aspects. Section \ref{sec:methodo} describes the
  Bayesian framework including the specific Markov Chain Monte Carlo
  (\textbf{MCMC}) algorithm, required to extend the work by
  \cite{Stephenson2004a}. Section \ref{sec:genProc} presents the
  simple and efficient algorithm to generate stochastically
  hydrological homogeneous regions. A sensitivity analysis is
  performed in section \ref{sec:sensAnal} to assess how quantile
  estimates and related uncertainties are influenced by the values of
  two parameters of the reversible jump Markov chains.  Section
  \ref{sec:simStud} compares the performance of each estimator on six
  representative case studies. The impact of the bias in the target
  site index flood estimation is analyzed in section
  \ref{sec:effBias}, while suggestions for building efficient pooling
  groups are presented in section \ref{sec:guidRegConf}.  Finally,
  some conclusions are drawn in section \ref{sec:disc}.

  \section{Methodology}
  \label{sec:methodo}

  In the Bayesian framework, the posterior distribution of parameters
  must be known to derive quantile estimates. The posterior
  distribution $\pi(\theta | x)$ is given by the Bayes
  Theorem~\citep{Bayes1763}:

  \begin{equation}
    \label{eq:bayesthm}
    \pi\left(\theta | x\right) = \frac{\pi\left(\theta\right) \pi\left(x
        ; \theta\right)}{\int_{\Theta}\pi\left(\theta\right) \pi\left(x
        ; \theta\right)d\theta } \propto \pi\left(\theta\right)
    \pi\left(x ; \theta\right)
  \end{equation}
  where $\theta$ is the vector of parameters of the distribution to be
  fitted, $\Theta$ is the parameter space. $\pi\left(x ;
    \theta\right)$ is the likelihood function, $x$ is the vector of
  observations and $\pi\left(\theta\right)$ is the prior distribution.

  In this study, as excesses over a high threshold are of interest,
  the likelihood function $\pi(x ; \theta)$ is related to the GPD -
  see equation~\eqref{eq:gpd}.

  \subsection{Prior Distribution}
  \label{subsec:priorDist}

  In this section, the methodology to elicit the prior distribution is
  presented. In this study, regional information is used to define the
  prior distribution. Furthermore, the prior is specific as it must
  account for a fixed shape parameter $\xi_\mathrm{Fix}$ with a non
  null probability $p_\xi$. Let $\Theta_0$ be a sub-space of the
  parameter space $\Theta$ of $\theta$. More precisely, $\Theta_0 =
  \left\{\theta \in \Theta : \xi = \xi_\mathrm{Fix} \right\}$. $p_\xi$
  is a hyper-parameter of the prior distribution. The approach is to
  construct a suitable prior distribution on $\Theta$; then, for
  $p_\xi$ fixed, to modify this prior to account for the probability
  of $\Theta_0$.

  For clarity purposes, the prior distribution is defined in two
  steps. First, an initial prior distribution
  $\pi_\mathrm{in}(\theta)$ defined on $\Theta$ is introduced. Second,
  a revised prior distribution $\pi(\theta)$ is derived from
  $\pi_\mathrm{in}(\theta)$ to attribute a non null probability to the
  $\Theta_0$ sub-sample.

  \subsubsection{Initial prior distribution}
  \label{sec:initPrior}

  As the proposed model is fully parametric, the initial prior
  distribution $\pi_\mathrm{in}(\theta)$ is a multivariate
  distribution entirely defined by its hyper-parameters. In our case
  study, the initial prior distribution corresponds to the one
  introduced by \citet{Ribatet2007a}. Consequently, the marginal prior
  distributions were supposed to be independent lognormal for both
  location and scale parameters and normal for the shape parameter.
  Thus,

  \begin{equation}
    \label{eq:prior}
    \pi_\mathrm{in}(\theta) \propto J \exp\left[ (\theta' - \gamma)^T
      \Sigma^{-1} (\theta' - \gamma) \right]
  \end{equation}
  where $\gamma, \Sigma$ are hyper-parameters, $\theta' = (\log \mu,
  \log \sigma, \xi)$ and $J$ is the Jacobian of the transformation
  from $\theta'$ to $\theta$, namely $J=1/\mu\sigma$.
  $\gamma=(\gamma_1, \gamma_2, \gamma_3)$ is the mean vector, $\Sigma$
  is the covariance matrix. As marginal priors are supposed to be
  independent, $\Sigma$ is a $3\times3$ diagonal matrix with diagonal
  elements $d_1, d_2, d_3$.

  Hyper-parameters are defined through the index flood concept, that
  is, all distributions are identical up to an at-site dependent
  constant. Consider all sites of a region except the target site -
  say the $j$-th site. A set of pseudo target site parameters can be
  computed:

  \begin{align}
    \label{eq:pseudoLoc}
    \tilde{\mu}_i &= C^{(j)} \mu_*^{(i)}\\
    \label{eq:pseudoScale}
    \tilde{\sigma}_i &= C^{(j)} \sigma_*^{(i)}\\
    \label{eq:pseudoShape}
    \tilde{\xi}_i &= \xi_*^{(i)}
  \end{align}
  for $i \neq j$, where $C^{(j)}$ is the target site index flood and
  $\mu_*^{(i)}, \sigma_*^{(i)}, \xi_*^{(i)}$ are respectively the
  location, scale and shape at-site parameter estimates from the
  rescaled sample - e.g. normalized by its respective index flood
  estimate. Under the hypothesis of the index flood concept,
  pseudo-parameters are expected to be distributed as parameters of
  the target site.

  Information from the target site sample can not be used to elicit
  the prior distribution. Thus, $C^{(j)}$ in
  equations~\eqref{eq:pseudoLoc} and~\eqref{eq:pseudoScale} must be
  estimated without use of the $j$-th sample site.

  In this case study, $C^{(j)}$ is estimated through a Generalized
  Linear Model (\textbf{GLM}) defined by:

  \begin{equation}
    \label{eq:glm}
    \begin{cases}
      \mathbb{E} \left[\log C^{(j)} \right] &= \nu, \qquad \nu =
      X\beta\\
      Var \left[ \log C^{(j)} \right] &= \phi V(\nu)
    \end{cases}
  \end{equation}
  where $X$ are basin characteristics (possibly log transformed),
  $\phi$ is the dispersion parameter, $V$ the variance function and
  $\nu$ is the linear predictor. \cite{McCullagh1989} give a
  comprehensive introduction to GLM\@. Other alternatives for modeling
  the target site index flood can be considered such as Generalized
  Additive Models~\citep{Wood2002}, Neural Networks~\citep{Shu2004} or
  Kriging~\citep{Merz2005}. However, the variance of $C^{(j)}$ should
  be estimated. Indeed, as $C^{(j)}$ is estimated without use of the
  target site data, uncertainties due to this estimation must be
  incorporated in the prior distribution.

  From these pseudo parameters, hyper-parameters can be computed:

  \begin{align}
    \label{eq:gamma1}
    \gamma_1 =& \frac{1}{N-1}\sum_{i\neq j} \log \tilde{\mu}^{(i)},
    &d_1 =& \frac{1}{N-1}\sum_{i \neq j} Var\left[ \log
      \tilde{\mu}^{(i)}
    \right] \\
    \label{eq:gamma2}
    \gamma_2 =& \frac{1}{N-1}\sum_{i\neq j} \log \tilde{\sigma}^{(i)},
    &d_2 =& \frac{1}{N-1}\sum_{i \neq j} Var\left[ \log
      \tilde{\sigma}^{(i)}
    \right] \\
    \label{eq:gamma3}
    \gamma_3 =& \frac{1}{N-1}\sum_{i\neq j} \tilde{\xi}^{(i)}, &d_3 =&
    \frac{1}{N-2}\sum_{i \neq j} \left(\tilde{\xi}^{(i)} - \gamma_3
    \right)^2
  \end{align}

  Under the independence assumption between $C^{(j)}$ and
  $\mu_*^{(i)}, \sigma_*^{(i)}$, the following relations hold:

  \begin{align}
    \label{eq:varPseudoLoc}
    Var\left[ \log \tilde{\mu}^{(i)} \right] &= Var\left[ \log C^{(j)}
    \right] + Var\left[ \log \mu_*^{(i)} \right]\\
    \label{eq:varPseudoScale}
    Var\left[ \log \tilde{\sigma}^{(i)} \right] &= Var\left[ \log
      C^{(j)} \right] + Var\left[ \log \sigma_*^{(i)} \right]
  \end{align}
  The independence assumption is not too restrictive as the target
  site index flood is estimated independently from $\mu_*^{(i)},
  \sigma_*^{(i)}$.

  Note that $Var\left[ \log \cdot_*^{(i)} \right]$ are estimated
  thanks to Fisher information and the delta method. Estimation of
  $\mathrm{Var}\left[ \log C^{(j)}\right]$ is a special case and
  depends on the method for estimating the at-site index flood.
  Nevertheless, it is always possible to carry out an estimation of
  this variance, at least through standard errors.

  \subsubsection{Revised prior distribution}
  \label{sec:newPrior}

  The initial prior distribution $\pi_\mathrm{in}(\theta)$ gives a
  null probability to the sub-sample $\Theta_0$. Thus, from this
  initial prior $\pi_\mathrm{in}(\theta)$, a revised prior
  $\pi(\theta)$ is constructed to attribute a non null probability to
  the $\Theta_0$ sub-sample. According to \cite{Stephenson2004a},
  $\pi(\theta)$ is defined as~:

  \begin{equation}
    \label{eq:priorDist}
    \pi(\theta) =
    \begin{cases}
      (1-p_\xi) \pi_\mathrm{in}(\theta) & \text{for } \theta \in
      \Theta \backslash
      \Theta_0\\
      p_\xi \pi_{\xi_\mathrm{Fix}}(\theta) & \text{for } \theta \in
      \Theta_0
    \end{cases}
  \end{equation}
  where $p_\xi \in [0,1]$ and with

  \begin{equation}
    \label{eq:normConst}
    \pi_{\xi_\mathrm{Fix}}(\theta) = \frac{\pi_\mathrm{in}(\mu, \sigma,
      \xi_\mathrm{Fix})} {\int_{\mu,\sigma} \pi_\mathrm{in}(\mu, \sigma,
      \xi_\mathrm{Fix}) d\mu d\sigma}
  \end{equation}
  for $\theta \in \Theta_0$. The integral in
  equation~\eqref{eq:normConst} can be easily evaluated by standard
  numerical integration methods.

  By construction, the new prior distribution $\pi(\theta)$ gives the
  required probability to the sub-space $\Theta_0$.
  \cite{Stephenson2004a} have already applied
  formulations~\eqref{eq:priorDist} and~\eqref{eq:normConst} with
  success for sea-level maxima and rainfall threshold exceedences.

  \subsection{Posterior Estimation}
  \label{subsec:postEst}

  As it is often the case in Bayesian analysis, the integral in
  equation~\eqref{eq:bayesthm} is insolvable analytically. MCMC
  techniques are used to overcome this problem. Yet, due to the
  duality of $\pi(\theta)$ distribution, standard Metropolis-Hastings
  \citep{Hastings1970} within Gibbs \citep{Geman1984} methods are not
  sufficient. Reversible jump techniques \citep{Green1995} are used to
  allow moves from the two dimensional space $\Theta_0$ to the three
  dimensional space $\Theta \backslash \Theta_0$ and vice-versa.

  The classical Bayesian analysis, on $\Theta \backslash \Theta_0$, is
  performed with Gibbs cycle over each component of $\theta$ using
  Metropolis-Hastings updates, with random walk
  proposals~\citep{Coles1996}.

  \cite{Stephenson2004a} extended this algorithm to incorporate the
  mass on the Gumbel/Exponential case. However, as our approach does
  not only focus on the $\xi_\mathrm{Fix} = 0$ case, a new algorithm
  must be implemented. To help understand the algorithmic
  developments, some details about the classical Metropolis-Hastings
  algorithm and the reversible jump case are reported in
  Appendix~\ref{sec:metro}.

  The proposed algorithm must deal with two dimensional changes: a
  change to $\Theta_0$ from $\Theta \backslash \Theta_0$ space and
  vice-versa. These two types of special moves must be defined
  cautiously. As inspired by~\cite{Stephenson2004a}, quantiles
  associated to a non exceedence probability $p$ are set to be equal
  at current state $\theta_t$ and proposal $\theta_\mathrm{prop}$, $p$
  being fixed.

  For a proposal move to $\Theta \backslash \Theta_0$ from $\Theta_0$,
  i.e., $\xi_t = \xi_\mathrm{Fix}$ and a proposal shape
  $\xi_\mathrm{prop} \neq \xi_\mathrm{Fix}$, the candidate move is to
  change $\theta_t = (\mu_t, \sigma_t, \xi_t)$ to
  $\theta_\mathrm{prop} = (\mu_\mathrm{prop}, \sigma_\mathrm{prop},
  \xi_\mathrm{prop})$ where

  \begin{subequations}
    \begin{align}
      \label{eq:proploc1}
      \mu_\mathrm{prop} &= \mu_t\\
      \label{eq:propscale1}
      \sigma_\mathrm{prop} &= \sigma_t \frac{\xi_\mathrm{prop}
        (y^{-\xi_t} - 1)}{\xi_t (y^{-\xi_\mathrm{prop}} - 1)}\\
      \label{eq:propshape1}
      \xi_\mathrm{prop} &\sim \mathcal{N}( \tilde{\xi}, s_\xi^2 )
    \end{align}
  \end{subequations}
  where $y=1-p$, $p$ being fixed, $\tilde{\xi}$ is taken to be the
  mode of the marginal distribution for $\xi$ when there is no mass on
  $\Theta_0$ \citep{Stephenson2004a}, and $s_\xi$ is the standard
  deviation selected to give good mixing properties to the chain. As
  it is usually the case with Metropolis-Hastings updates, this move
  is accepted with probability $\min(1, \Delta)$ with

  \begin{equation}
    \label{eq:delta}
    \Delta = \frac{\pi(\mu_\mathrm{prop}, \sigma_\mathrm{prop},
      \xi_\mathrm{prop} | x)}{\pi(\mu_t, \sigma_t, 
      \xi_\mathrm{Fix} | x)} \frac{p_\xi}{1-p_\xi} \left[
      \phi(\xi_\mathrm{prop}; \tilde{\xi}, s_\xi^2 )
      J_{\xi_\mathrm{Fix}} (\xi_\mathrm{prop}) \right]^{-1} 
  \end{equation}
  where $\phi(\cdot; m, s^2 )$ denotes the density function of the
  Normal distribution with mean $m$ and variance $s^2$, and
  $J_{\xi_\mathrm{Fix}}$ is the Jacobian of the parameter
  transformation for quantile matching, that is:

  \begin{equation}
    \label{eq:jacobian}
    J_{\xi_\mathrm{Fix}}(\xi) = \frac{\xi_\mathrm{Fix}}{\xi}
    \frac{y^{-\xi} - 1}{y^{-\xi_\mathrm{Fix}} - 1}
  \end{equation}
  If the move is accepted, then $\theta_{t+1} = (\mu_\mathrm{prop},
  \sigma_\mathrm{prop}, \xi_\mathrm{prop})$, else $\theta_{t+1} =
  \theta_t$.

  For a proposal move to $\Theta_0$ from $\Theta \backslash \Theta_0$,
  i.e., $\xi_t \neq \xi_\mathrm{Fix}$ and a proposal shape
  $\xi_\mathrm{prop} = \xi_\mathrm{Fix}$, the proposal is to change
  $\theta_t = (\mu_t, \sigma_t, \xi_t)$ to $\theta_\mathrm{prop} =
  (\mu_\mathrm{prop}, \sigma_\mathrm{prop}, \xi_\mathrm{prop})$ where

  \begin{subequations}
    \begin{align}
      \label{eq:proploc2}
      \mu_\mathrm{prop} &= \mu_t\\
      \label{eq:propscale2}
      \sigma_\mathrm{prop} &= \sigma_t \frac{\xi_\mathrm{prop}
        (y^{-\xi_t} - 1)}{\xi_t (y^{-\xi_\mathrm{prop}} - 1)}\\
      \label{eq:propshape2}
      \xi_\mathrm{prop} &= \xi_\mathrm{Fix}
    \end{align}
  \end{subequations}
  This move is accepted with probability $\min(1, \Delta)$ where

\begin{equation}
  \Delta = \frac{\pi(\mu_\mathrm{prop}, \sigma_\mathrm{prop},
    \xi_\mathrm{Fix} | x)}{\pi(\mu_t, \sigma_t, \xi_t | x)} \frac{1 -
    p_\xi}{p_\xi} \phi(\xi_t; \tilde{\xi}, s_\xi^2 )
  J_{\xi_\mathrm{Fix}} (\xi_t)
\end{equation}
If the move is accepted, then $\theta_{t+1} = (\mu_\mathrm{prop},
\sigma_\mathrm{prop}, \xi_\mathrm{prop})$ else $\theta_{t+1} =
\theta_t$.

Obviously, special moves introduced in this study are not the only
conceivable ones. Other reversible jumps can be explored - see for
example~\cite{Stephenson2004a}. However, for this application, the
proposed moves seem to be particularly well suited. Indeed, a
preliminary study shows that the location parameter was well estimated
by a regional Bayesian approach. Thus, a special move which only
affects the shape and scale parameters should be consistent.

\section{Generation Procedure}
\label{sec:genProc}

In this section, the procedure implemented to generate stochastic
homogeneous regions is described. The idea consists in generating
sample points in a neighborhood of the L-moment space (Mean, L-CV,
L-Skewness). The generation procedure can be summarized as follows:
\begin{enumerate}
\item Set the center of the neighborhood i.e. $(l_{1,R}, \tau_{R},
  \tau_{3,R})$ or equivalently parameters of the regional distribution
  $(\mu_R, \sigma_R, \xi_R)$;
\item Generate $N$ points $(l_{1,i}, \tau_{i}, \tau_{3,i})$ uniformly
  in the sphere $\mathcal{B}\left((l_{1,R}, \tau_{R}, \tau_{3,R}) ;
    \varepsilon\right)$;
\item Generate $N$ index floods $C$ using the scaling model
  parametrization:

  \begin{equation}
    \label{eq:GLMparam}
    C = \alpha Area^\beta
  \end{equation}
  Catchment areas are defined as realizations of a lognormal random
  variable.
\item For each $(l_{1,i}, \tau_{i}, \tau_{3,i})$, compute adimensional
  parameters by:

  \begin{subequations}
    \begin{align}
      \xi_i^* &= \frac{3\tau_{3,i} - 1}{1+\tau_{3,i}}\\
      \sigma_i^* &= (\xi_i^*-1)(\xi_i^*-2)l_{1,i}\tau_i\\
      \mu_i^* &= l_{1,i} - \frac{\sigma_i^*}{1-\xi_i^*}
    \end{align}
  \end{subequations}
\item Then, compute at-site parameters from:

  \begin{subequations}
    \begin{align}
      \xi_i &= \xi_i^*\\
      \sigma_i &= C_i \sigma_i^*\\
      \mu_i &= C_i \mu_i^*
    \end{align}
  \end{subequations}
\item Simulate samples from a GPD with parameters $(\mu_i, \sigma_i,
  \xi_i)$.
\end{enumerate}

As a GLM is used to elicit the prior distribution, the scaling
model~\eqref{eq:GLMparam} must be altered to avoid giving an advantage
to the Bayesian approaches over the index flood model. For this
purpose, a noise in relation~\eqref{eq:GLMparam} at step 3 is
introduced. Thus, areas are altered by adding uniform random variables
varying in $\left(-0.5 \times Area, 0.5 \times Area\right)$.

\begin{figure}
  \centering
  \includegraphics[angle=-90,width=0.5\textwidth]{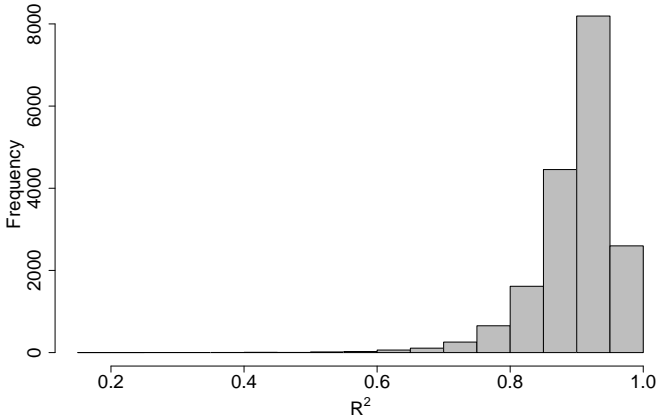}
  \caption{Histogram of the coefficient of determination for the
    regressive model~\eqref{eq:glm}. Application of
    section~\ref{sec:simStud}.}
  \label{fig:histR2}
\end{figure}

This distortion is necessary to ensure that the regressive model is
not too competitive and is consistent with observations. Indeed large
deviations to the area-index flood relationship are often encountered
in practice. In the following applications, $\alpha = 0.12,
\beta=1.01$ and $Area \sim \mathcal{LN}(4.8, 1)$. These values arise
from a previous study on a French data set~\citep{Ribatet2007a} and
ensure realistic magnitudes. For the application of
section~\ref{sec:simStud}, the coefficients of determination for the
regressive model~\eqref{eq:glm} varies from 0.20 to 0.99, with a mean
value of 0.89. The histogram of these coefficients of determination is
presented in Figure~\ref{fig:histR2}. The radius $\varepsilon$ in the
generation algorithm is set to 0.04. This value is chosen to reflect
variability met in practice while preserving a low dispersion around
the regional distribution. The $\varepsilon$ value primarily impacts
the proportions of regions satisfying $H_1 < 1$. For specific
applications, regions with a heterogeneity statistic $H_1$ such as
$H_1 > 1$ may be discarded.

\section{Sensitivity Analysis}
\label{sec:sensAnal}

In this section, a sensitivity analysis for the algorithm introduced
in section~\ref{subsec:postEst} is carried out. The primary goal is to
check if results are not too impacted by the choice of the two
user-selectable parameters $p_\xi$ and $\xi_\mathrm{Fix}$. For this
purpose, the effect of both $p_\xi$ and $\xi_\mathrm{Fix}$ values on
estimates and credibility intervals is examined. For this sensitivity
analysis, the parameters of the regional distribution is set to be
(0.64, 0.48, 0.26). The regions have 20 sites with a sample size of
70. For the whole sensitivity analysis, 10 000 regions were generated.
The target site has a sample size of 10. We concentrate on estimates
at sites with very few data, to exhibit the main differences in the
most restricting configuration. Other configurations were found to
demonstrate features similar to Fig.~\ref{fig:senspMassProbQ} and
Fig.~\ref{fig:uncertShape}.

\subsection{Effect of $p_\xi$}
\label{subsec:effpxi}

\begin{figure*}
  \centering
  \includegraphics[width=\textwidth]{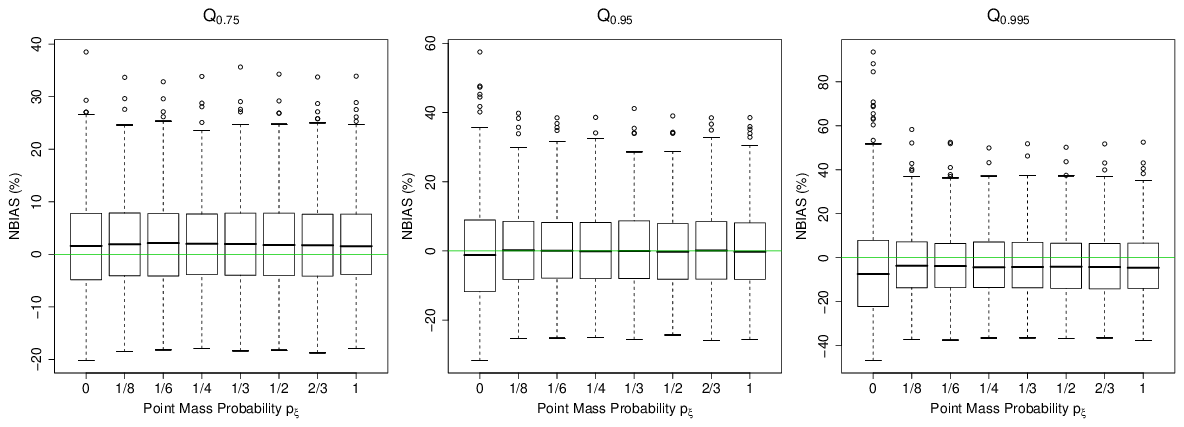}
  \caption{Effect of $p_\xi$ value on quantile estimation with non
    exceedence probabilities 0.75, 0.95 and 0.995. Sample size 10.
    $\xi_\mathrm{Fix} = 0.26$.}
  \label{fig:senspMassProbQ}
\end{figure*}

The evolution of the normalized biases (expressed in percent) for
return levels with non exceedence probabilities 0.75, 0.95 and 0.995
associated to several $p_\xi$ values are depicted in
Fig.~\ref{fig:senspMassProbQ}. Each boxplot is obtained from at-site
estimates computed on more than 365 stochastic homogeneous regions.
The case $p_\xi = 0$ corresponds to a classical Bayesian approach free
from any point mass. In addition, to analyze only the effect of the
parameter $p_\xi$, $\xi_\mathrm{Fix}$ is temporarily fixed to be equal
to the theoretical regional shape parameter.

From Fig.~\ref{fig:senspMassProbQ}, the quantile estimates
distribution seems to be stationary, provided that $p_\xi > 0$.
Introducing a point mass does not impact $Q_{0.75}$ estimates, whereas
significant reduction in median biases and scatter of estimates is
noticeable for more extremal quantiles.

\begin{figure*}
  \centering
  \includegraphics[angle=-90,width=\textwidth]{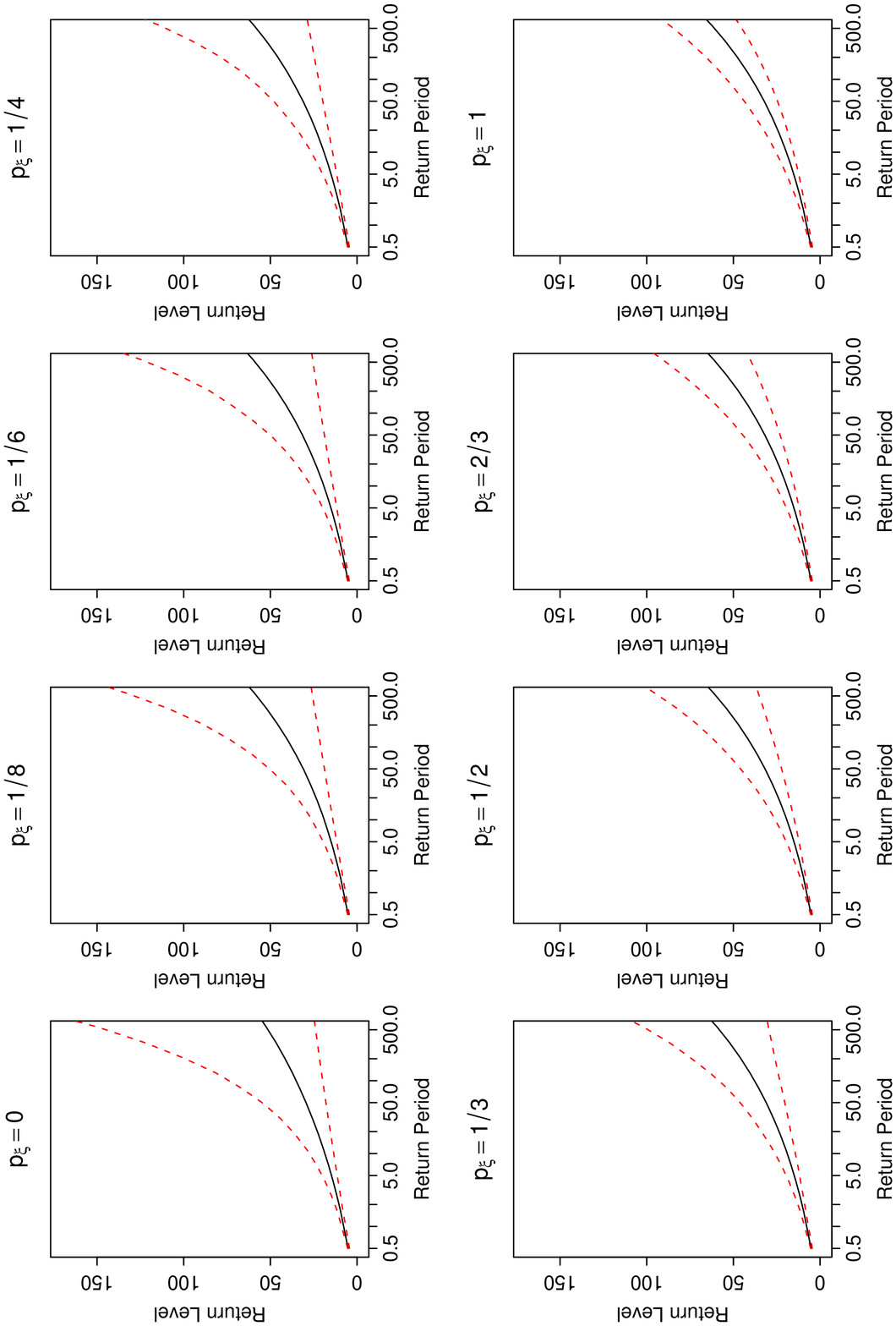}
  \caption{Effect of $p_\xi$ value on 90\% posterior credibility
    interval. Sample Size 10.}
  \label{fig:senspMassProbCI}
\end{figure*}

Fig.~\ref{fig:senspMassProbCI} shows the posterior distributions of
return levels and 90\% posterior credibility intervals for several
$p_\xi$ values.

It is clear that credibility intervals are sensitive to the $p_\xi$
value.  This result is consistent as more and more proposals in the
MCMC simulation belong to $\Theta_0$ as $p_\xi$ increases. Thus, by
construction, the Markov chain is less variable. As denoted by
\cite{Stephenson2004a}, the special case $p_\xi = 1$ is particular as
uncertainty in the shape parameter is not considered.  In that case,
credibility intervals could be falsely narrow.

\subsection{Effect of $\xi_\mathrm{Fix}$}
\label{subsec:effxifix}

\begin{table}
  \caption{Posterior proportions (in percent) of events $\left\{\theta
      \in \Theta_0\right\}$ for different values of $p_\xi$ and 
    $\xi_\mathrm{Fix}$. Target Sample Size 60.}
  \label{tab:sensXiFixProp}
  \centering
  \begin{tabular}{rrrrrrrrr}
    \hline
    \multicolumn{2}{c}{$\xi_\mathrm{Fix}$ features} &
    \multicolumn{7}{c}{$p_\xi$ values}\\
    \cline{1-2}  \cline{4-9}
    $R_\mathrm{Shape}$ & $D_\mathrm{Shape}$ && 1/8 & 1/6 & 1/4 & 1/3 &
    1/2 & 2/3\\ 
    \hline
    -0.50 & 2e$^{-5}$  && 0.00  &  0.03 & 0.00 & 0.00 & 0.05 & 0.00 \\
    0.00 & 0.06  && 10.07 & 14.55 & 17.27 & 21.99 & 41.53 & 61.84 \\ 
    0.50 & 0.70  && 38.88 & 46.94 & 59.96 & 67.42 & 81.88 & 92.17 \\
    0.83 & 1.00  && 46.21 & 57.33 & 67.53 & 76.08 & 85.33 & 92.20 \\
    1.00 & 0.87  && 48.24 & 55.14 & 68.90 & 76.16 & 86.05 & 91.85 \\
    1.50 & 0.41  && 32.72 & 45.61 & 54.62 & 66.18 & 82.11 & 89.90 \\
    2.00 & 0.10  && 22.95 & 22.83 & 35.06 & 49.82 & 57.86 & 81.92 \\
    2.50 & 0.01  && 13.93 & 7.04  &  9.86 & 36.21 & 38.89 & 42.28 \\
    \hline
  \end{tabular}
\end{table}

It is important to analyze the influence of the choice of
$\xi_\mathrm{Fix}$ on the simulated Markov chains; and thus, its
impact on estimations.  Indeed, when specifying an unreasonable
$\xi_\mathrm{Fix}$ value, the estimations must not differ
significantly from the conventional Bayesian ones. For this purpose,
Tab.~\ref{tab:sensXiFixProp} displays the posterior proportions of
events $\left\{\theta \in \Theta_0\right\}$ for several
$\xi_\mathrm{Fix}$ and $p_\xi$ values. This table is obtained with a
target site sample size of 60. For each specified $\xi_\mathrm{Fix}$
value, two features are computed to measure the relevance of the
$\xi_\mathrm{Fix}$ value: (a) $R_\mathrm{Shape}$ the ratio of
$\xi_\mathrm{Fix}$ to the true shape parameter ; and (b)
$D_\mathrm{Shape}$ the ratio of the marginal posterior density from a
conventional Bayesian analysis evaluated in $\xi_\mathrm{Fix}$ and
$\tilde{\xi}$.

$R_\mathrm{Shape}$ characterizes how much the point Mass differs from
the true value. $D_\mathrm{Shape}$ quantifies the distance of
$\xi_\mathrm{Fix}$ from the estimator of the shape parameter proposed
by~\cite{Ribatet2007a}. Thus, from these two statistics, consistency
of the posterior proportions with deviations from theoretical and
empirical values can be analyzed.

\begin{figure}
  \centering
  \includegraphics[angle=-90,width=0.5\textwidth]{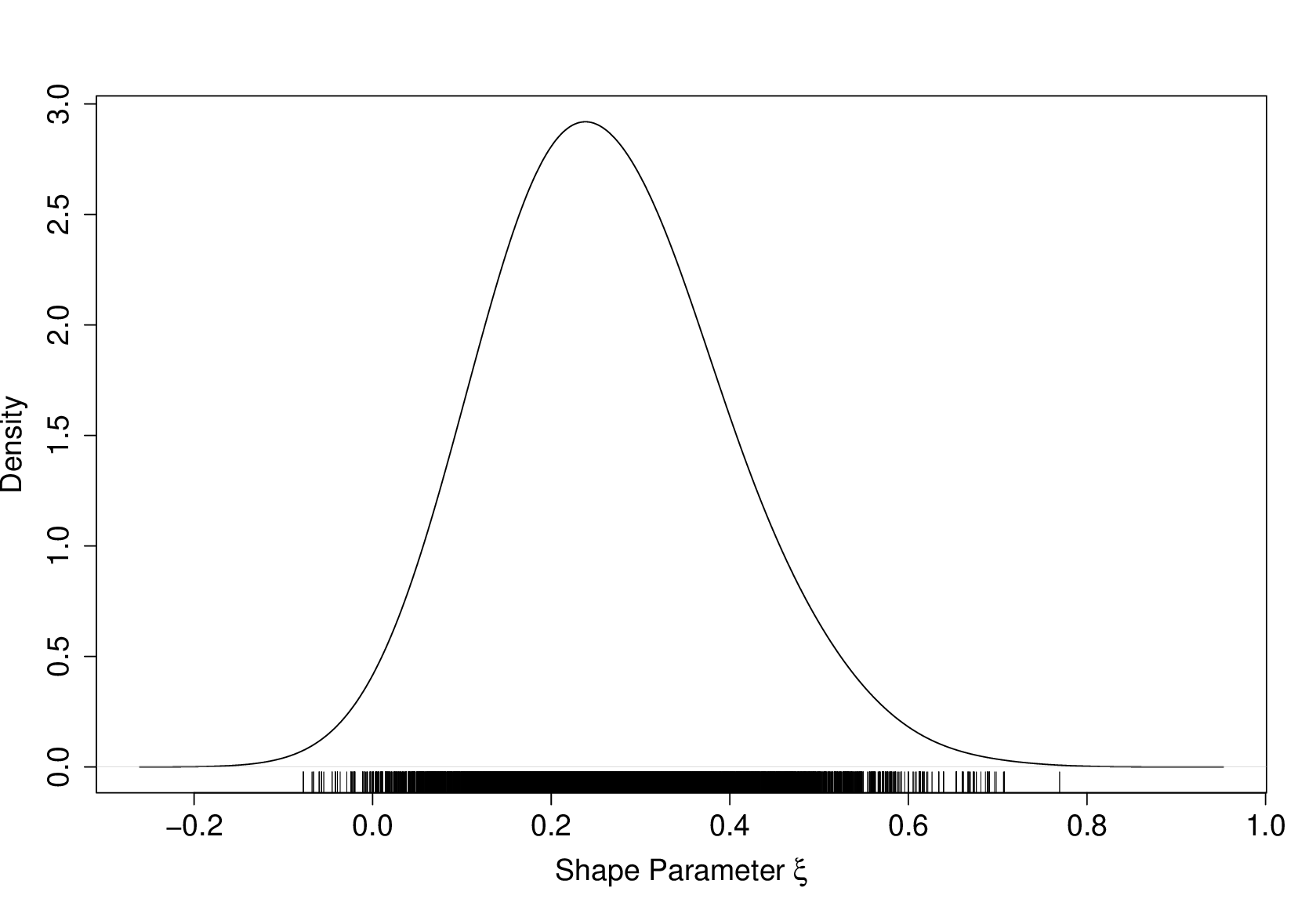}
  \caption{Posterior marginal density for the shape parameter.}
  \label{fig:uncertShape}
\end{figure}

The results in Tab.~\ref{tab:sensXiFixProp} show that values of
$\xi_\mathrm{Fix}$ that are not consistent with the data imply low
proportions of state in $\Theta_0$. Thus, for such values, the
proposed model is quite similar to a conventional Bayesian analysis.
However, for two different values of $\xi_\mathrm{Fix}$
($R_\mathrm{Shape}$ equal to 0.83 and 1), the posterior proportions
are quite equivalent. This emphasizes the large uncertainty on the
shape parameter estimation for small sample sizes. Uncertainty on the
shape parameter estimation is also corroborated by the posterior
marginal distribution of a conventional Bayesian analysis - see
Fig.~\ref{fig:uncertShape}.

As noticed above, these results are obtained with a target site sample
size of 60. This particular sample size was selected as it is the most
illustrative case. However, the posterior proportions are quite
similar when dealing with other target site sample sizes - even if for
very small sample sizes, this is less noticeable.

\section{Simulation Study}
\label{sec:simStud}

In this section, performance of three different estimators are
analyzed: a conventional Bayesian estimator ($\mathbf{BAY}$)
introduced by~\cite{Ribatet2007a}, the proposed estimator based on
reversible jumps ($\mathbf{REV}$) and the index flood estimator
($\mathbf{IFL}$). In particular, the $BAY$ estimator is related to the
initial prior distribution defined in Section~\ref{sec:initPrior}.
Thus, the $BAY$ estimator is identical to the $REV$ approach with
$p_\xi = 0$.

For the proposed estimator, the point Mass probability $p_\xi$ was set
to be a function of the $H_1$ statistic of \cite{Hosking1997}; that
is:

\begin{equation}
  \label{eq:defpxi}
  p_\xi = \frac{\exp (-H_1)} {1 + \exp (-H_1 )}
\end{equation}

For this parametrization, necessary requirements are satisfied; i.e.
$p_\xi \rightarrow 0$ when $H_1 \rightarrow +\infty$ and $p_\xi
\rightarrow 1$ when $H_1 \rightarrow -\infty$. Moreover, for $H_1 =
0$, $p_\xi = 0.5$ which corresponds to the estimator introduced by
\cite{Stephenson2004a}.  Note that $p_\xi$ in Eq.~\eqref{eq:defpxi} is
defined with the negative inverse of the so called \emph{logit}
function.

Thus, for this choice, as underlined by the sensitivity analysis,
credibility intervals are related to the degree of confidence of the
point Mass $\xi_\mathrm{Fix}$ to be the true shape parameter and
implicitly to the level of homogeneity of the regions.

In addition, the non exceedence probability $p$ used for quantiles
matching in our algorithm (see Section~\ref{subsec:postEst}) is equal
to $1 - 1/2n$, where $n$ is the target site sample size. This last
point guarantees that quantiles associated with non exceedence
probability $1 - 1/2n$ for both proposal and current state of the
Markov chain are identical. Other choices for $p$ are arguable. Here,
we introduce a quantile matching equation for a value closely related
to the scale parameter and for which uncertainties are not too large.


\begin{table}
  \caption{Characteristics of the sixth case studies. The target site
    is omitted in the couple $(n_\mathrm{Site}, n_\mathrm{Size})$ and
    has a sample size of: 10, 25 and 40.}
  \label{tab:caseStudConf}
  \centering
  \begin{tabular}{ccccc}
    \hline
    & $(\mu_R, \sigma_R, \xi_R)$ & $N_\mathrm{Site}$ &
    $(n_\mathrm{Site}, n_\mathrm{Size})$ & $N_\mathrm{Events}$\\
    \hline
    Conf1 & (0.64, 0.48, 0.26) & 10 & $(9, 50)$ & 450\\
    Conf2 & (0.64, 0.48, 0.26) & 20 & $(9, 30)\times(10, 18)$ & 450\\ 
    Conf3 & (0.64, 0.48, 0.26) & 15 & $(14, 50)$ & 700\\
    Conf4 & (0.66, 0.48, 0.08) & 10 & $(9, 50)$ & 450\\
    Conf5 & (0.66, 0.48, 0.08) & 20 & $(9, 30)\times(10, 18)$ & 450\\
    Conf6 & (0.66, 0.48, 0.08) & 15 & $(14, 50)$ & 700\\
    \hline
  \end{tabular}
\end{table}

The analysis was performed on six different case studies summarized in
Tab.~\ref{tab:caseStudConf}. The configurations differ by the way
information is distributed in space; that is, (a) ``small regions''
with well instrumented but few sites ($Conf1$ and $Conf4$); (b)
``large regions'' with less instrumented and numerous sites ($Conf2$
and $Conf5$) and (c) ``medium regions'' with well instrumented sites
and an intermediate number of gauging stations.  $Conf1$ (resp.
$Conf2$, $Conf3$) correspond to $Conf4$ (resp. $Conf5$, $Conf6$) apart
from the $(\mu_R, \sigma_R, \xi_R)$ values. The target site sample
size takes the values in 10, 25 and 40. 1000 regions were generated
for each configuration. Markov chains of length 15 000 were generated.
To ensure good mixing properties for all simulated Markov chains, an
automated trial and error process was used to define proposal standard
deviations of the MCMC algorithm. Furthermore, the first $2000$
iterations were discarded to ensure that the equilibrium was reached.

The performance of each estimator is assessed through the three
following statistics:
\begin{eqnarray}
  NBIAS &=& \frac{1}{k}\sum_{i=1}^k \frac{\hat{Q_i} - Q_i}{Q_i}\\
  SD &=& \sqrt{\frac{1}{k-1}\sum_{i=1}^k \left(\frac{\hat{Q_i} -
        Q_i}{Q_i} - NBIAS\right)^2}\\
  NMSE &=& \frac{1}{k}\sum_{i=1}^k \left(\frac{\hat{Q_i} -
      Q_i}{Q_i}\right)^2  
\end{eqnarray}
where $\hat{Q_i}$ is the estimate of the theoretical value $Q_i$ and
$k$ is the total number of theoretical values.

\subsection{$BAY$ vs. $IFL$ Approach}
\label{subsec:BAYvsIFL}


\begin{table*}
  \caption{Performance of $BAY$ and $IFL$ estimators for quantile
    $Q_{0.75}, Q_{0.95}$ and $Q_{0.995}$. Target site sample size:
    10.}
  \label{tab:perfSize10}
  \centering
  \begin{tabular}{rrrrrrrrrrrr}
    \hline
    \multirow{2}*{Model} & \multicolumn{3}{c}{$Q_{0.75}$} & &
    \multicolumn{3}{c}{$Q_{0.95}$} & &
    \multicolumn{3}{c}{$Q_{0.995}$}\\
    \cline{2-4} \cline{6-8} \cline{10-12}
    & $NBIAS$ & $SD$ & $NMSE$ && $NBIAS$ & $SD$ & $NMSE$ && $NBIAS$ &
    $SD$ & $NMSE$ \\ 
    \hline
    & \multicolumn{11}{c}{Conf1}\\
    $BAY$ & 0.015 & 0.123 & 0.015 && 0.001 & 0.187 & 0.035 &&
    $-$0.006 & 0.318 & 0.101 \\ 
    $IFL$ & 0.037 & 0.189 & 0.037 && 0.025 & 0.195 & 0.038 &&
    $-$0.004 & 0.230 & 0.053 \\ 
    & \multicolumn{11}{c}{Conf2}\\
    $BAY$ & 0.019 & 0.122 & 0.015 && 0.030 & 0.249 & 0.063 && 0.110
    & 0.561 & 0.326 \\ 
    $IFL$ & 0.041 & 0.183 & 0.035 && 0.025 & 0.191 & 0.037 &&
    $-$0.022 & 0.221 & 0.049 \\ 
    & \multicolumn{11}{c}{Conf3}\\
    $BAY$ & 0.019 & 0.110 & 0.012 && 0.006 & 0.174 & 0.030 &&
    $-$0.003 & 0.292 & 0.085 \\ 
    $IFL$ & 0.035 & 0.188 & 0.037 && 0.025 & 0.195 & 0.039 &&
    $-$0.002 & 0.222 & 0.049 \\ 
    & \multicolumn{11}{c}{Conf4}\\
    $BAY$ & 0.009 & 0.104 & 0.011 && $-$0.007 & 0.149 & 0.022 &&
    $-$0.021 & 0.233 & 0.054 \\ 
    $IFL$ & 0.023 & 0.157 & 0.025 && 0.022 & 0.163 & 0.027 && 0.022
    & 0.192 & 0.037 \\ 
    & \multicolumn{11}{c}{Conf5}\\
    $BAY$ & 0.018 & 0.109 & 0.012 && 0.012 & 0.193 & 0.037 && 0.033
    & 0.378 & 0.144 \\ 
    $IFL$ & 0.036 & 0.168 & 0.029 && 0.033 & 0.173 & 0.031 && 0.024
    & 0.197 & 0.039 \\ 
    & \multicolumn{11}{c}{Conf6}\\
    $BAY$ & 0.024 & 0.103 & 0.011 && 0.001 & 0.151 & 0.023 &&
    $-$0.038 & 0.222 & 0.050 \\ 
    $IFL$ & 0.028 & 0.168 & 0.029 && 0.028 & 0.177 & 0.032 && 0.028
    & 0.202 & 0.042 \\ 
    \hline
  \end{tabular}
\end{table*}

Table~\ref{tab:perfSize10} shows that, for a small target site sample
size and quantiles $Q_{0.75}$ and $Q_{0.95}$, the $BAY$ approach is
more competitive than the $IFL$ one. Indeed, the three $BAY$
statistics ($NBIAS$, $SD$, $NMSE$) are smaller than the ones related
to $IFL$. However, for $Conf2$ and $Conf5$, $IFL$ $Q_{0.95}$ estimates
are more competitive. These two case studies correspond to the same
configuration - i.e. numerous sites with short records.  $IFL$
estimates for $Q_{0.995}$ are always more accurate than $BAY$ for all
configurations.

These results indicate that the relative performance of $BAY$ compared
to $IFL$ depends on the pooling group. Thus, for the $BAY$ approach
and quantiles $Q_{0.75}$ and $Q_{0.95}$, it seems preferable to work
with less gauging stations but which have larger data series,
independently of the target site sample size. The sensitivity to the
configuration of the sites and the availability of long time series is
a drawback for the application of this Bayesian approach.

These conclusions obtained on stochastic regions are in line with a
previous analysis on a French data set~\citep{Ribatet2007a}. The $BAY$
approach is suited to work with ``small'' or ``medium'' regions and
well instrumented gauging stations. In addition, this approach is
accurate for ``reasonable'' quantile estimation -- see the bad
performance of $BAY$ for $Q_{0.995}$ in table~\ref{tab:perfSize10}.

However, the white noise introduced in the generation procedure is
independent of the target site sample size. It only regards both
Bayesian approaches. Thus, the performance of the $BAY$ estimator for
large sample sizes may be too impacted. Indeed, while the $IFL$
estimation procedure is not altered, both Bayesian approaches must
deal with artificially generated biases.

The main idea for the $REV$ approach is to combine the good
performance of the $BAY$ estimator for ``reasonable'' quantiles and
the efficiency of the $IFL$ approach for larger quantiles.

\subsection{$BAY$ vs. $REV$ Approach}
\label{sec:BAYvsREV}


\begin{table*}
  \caption{Performance of $BAY$ and $REV$ estimators for quantile
    $Q_{0.75}, Q_{0.95}$ and $Q_{0.995}$. Target site sample size:
    10.}
  \label{tab:perfSize10bis}
  \centering
  \begin{tabular}{rrrrrrrrrrrr}
    \hline
    \multirow{2}*{Model} & \multicolumn{3}{c}{$Q_{0.75}$} & &
    \multicolumn{3}{c}{$Q_{0.95}$} & &
    \multicolumn{3}{c}{$Q_{0.995}$}\\
    \cline{2-4} \cline{6-8} \cline{10-12}
    & $NBIAS$ & $SD$ & $NMSE$ && $NBIAS$ & $SD$ & $NMSE$ && $NBIAS$ &
    $SD$ & $NMSE$ \\ 
    \hline
    & \multicolumn{11}{c}{Conf1}\\
    $BAY$ & 0.015 & 0.123 & 0.015 && 0.001 & 0.187 & 0.035 && $-$0.006
    & 0.318 & 0.101 \\ 
    $REV$ & 0.011 & 0.119 & 0.014 && $-$0.012 & 0.159 & 0.026 &&
    $-$0.046 & 0.213 & 0.047 \\ 
    & \multicolumn{11}{c}{Conf2}\\
    $BAY$ & 0.019 & 0.122 & 0.015 && 0.030 & 0.249 & 0.063 && 0.110 &
    0.561 & 0.326 \\ 
    $REV$ & 0.005 & 0.105 & 0.011 && $-$0.026 & 0.154 & 0.024 &&
    $-$0.066 & 0.269 & 0.077 \\ 
    & \multicolumn{11}{c}{Conf3}\\
    $BAY$ & 0.019 & 0.110 & 0.012 && 0.006 & 0.174 & 0.030 && $-$0.003
    & 0.292 & 0.085 \\ 
    $REV$ & 0.014 & 0.103 & 0.011 && $-$0.008 & 0.139 & 0.019 &&
    $-$0.042 & 0.185 & 0.036 \\ 
    & \multicolumn{11}{c}{Conf4}\\
    $BAY$ & 0.009 & 0.104 & 0.011 && $-$0.007 & 0.149 & 0.022 &&
    $-$0.021 & 0.233 & 0.054 \\ 
    $REV$ & 0.010 & 0.102 & 0.011 && 0.002 & 0.136 & 0.018 && $-$0.001
    & 0.182 & 0.033 \\ 
    & \multicolumn{11}{c}{Conf5}\\
    $BAY$ & 0.018 & 0.109 & 0.012 && 0.012 & 0.193 & 0.037 && 0.033 &
    0.378 & 0.144 \\ 
    $REV$ & 0.013 & 0.097 & 0.010 && 0.000 & 0.126 & 0.016 && $-$0.014
    & 0.171 & 0.030 \\ 
    & \multicolumn{11}{c}{Conf6}\\
    $BAY$ & 0.024 & 0.103 & 0.011 && 0.001 & 0.151 & 0.023 && $-$0.038
    & 0.222 & 0.050 \\ 
    $REV$ & 0.031 & 0.099 & 0.011 && 0.033 & 0.133 & 0.019 && 0.034 &
    0.174 & 0.032 \\ 
    \hline
  \end{tabular}
\end{table*}

The comparison of the two Bayesian estimators is summarized in
Tab.~\ref{tab:perfSize10bis}. $REV$ leads to more accurate estimated
quantiles, in particular for $Q_{0.95}$ and $Q_{0.995}$. This last
point confirms the benefits of using a regional shape parameter
through a reversible jump approach.

By construction of the algorithm described in
section~\ref{subsec:postEst}, Markov chains generated from the $REV$
approach are less variable than the ones generated from the $BAY$
model. Thus, $REV$ is associated to smaller standard deviation than
$BAY$ whatever the configuration is (Table~\ref{tab:perfSize10bis}).
Moreover, if the regional fixed shape parameter $\xi_\mathrm{Fix}$ is
suited, $REV$ should have the same biases than $BAY$. Thereby, the
$REV$ estimator always leads to a smaller $NMSE$.

\subsection{Global Comparison}
\label{subsec:globEff}

\begin{figure*}
  \centering
  \includegraphics[angle=-90,width=\textwidth]{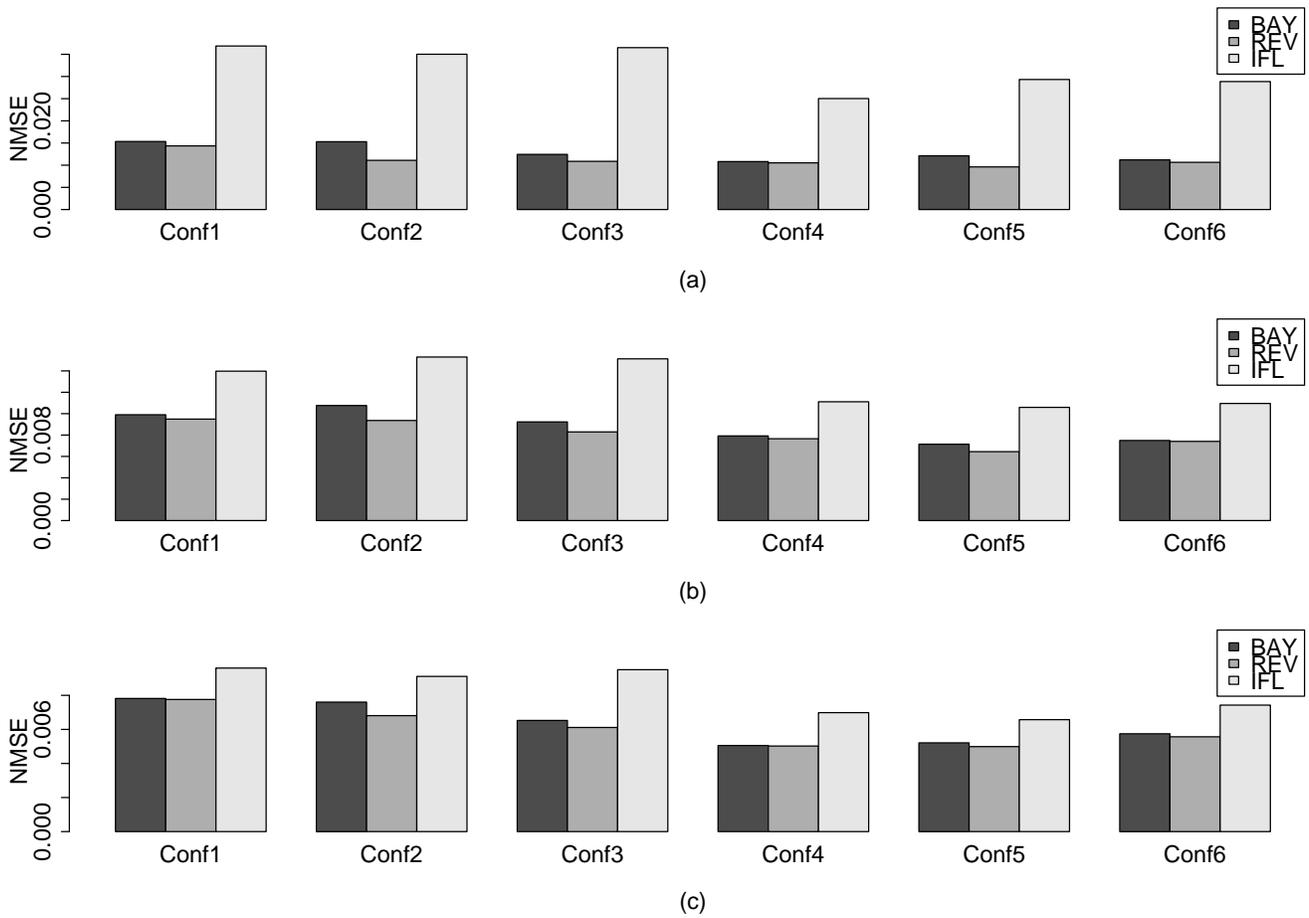}
  \caption{Evolution of the $NMSE$ for quantile $Q_{0.75}$ in function
    of the region configuration. Target site sample size: (a) 10, (b)
    25 and (c) 40.}
  \label{fig:globMSEQ75}
\end{figure*}

\begin{figure*}
  \centering
  \includegraphics[angle=-90,width=\textwidth]{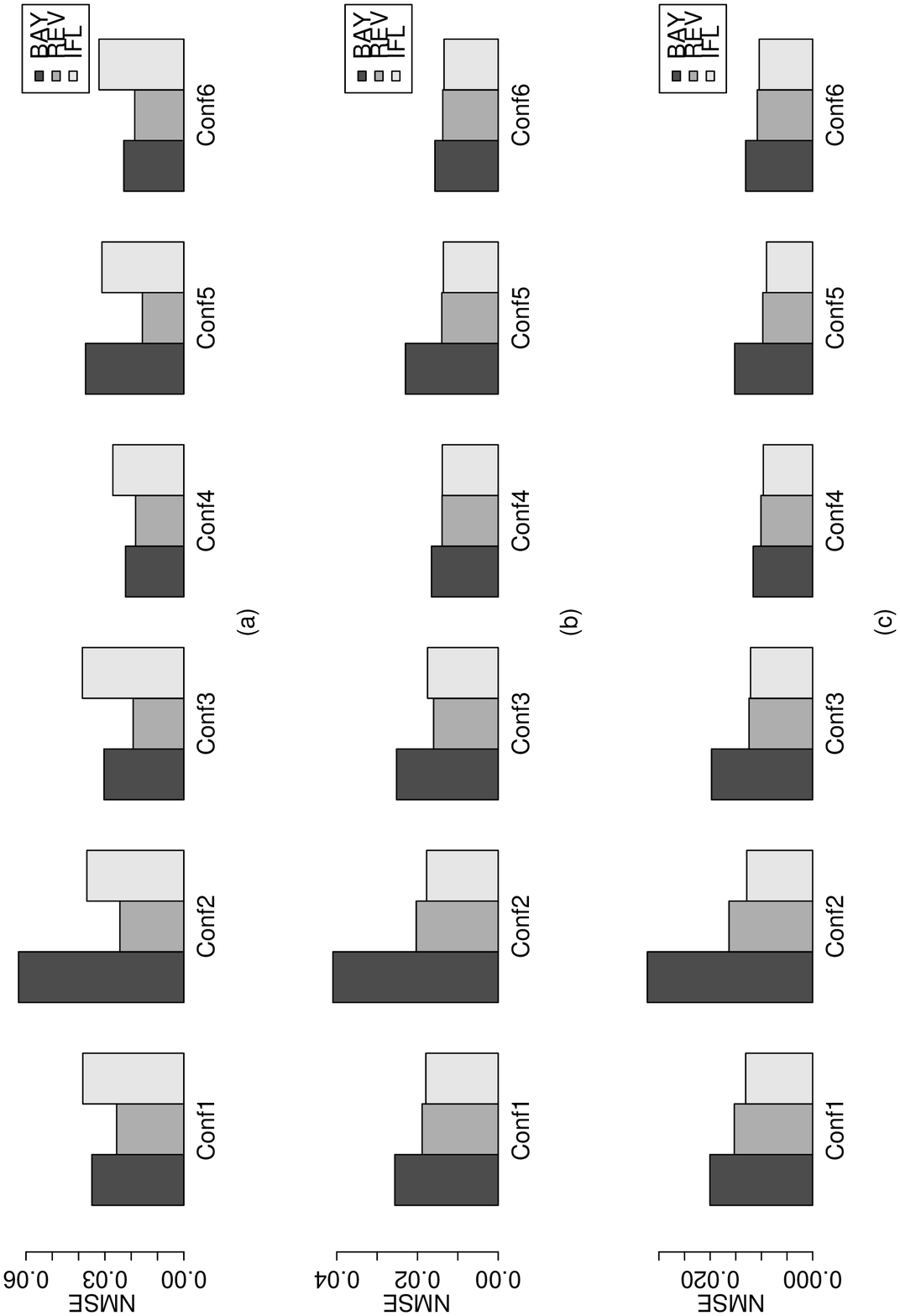}
  \caption{Evolution of the $NMSE$ for quantile $Q_{0.95}$ in function
    of the region configuration. Target site sample size: (a) 10, (b)
    25 and (c) 40.}
  \label{fig:globMSEQ95}
\end{figure*}

\begin{figure*}
  \centering
  \includegraphics[angle=-90,width=\textwidth]{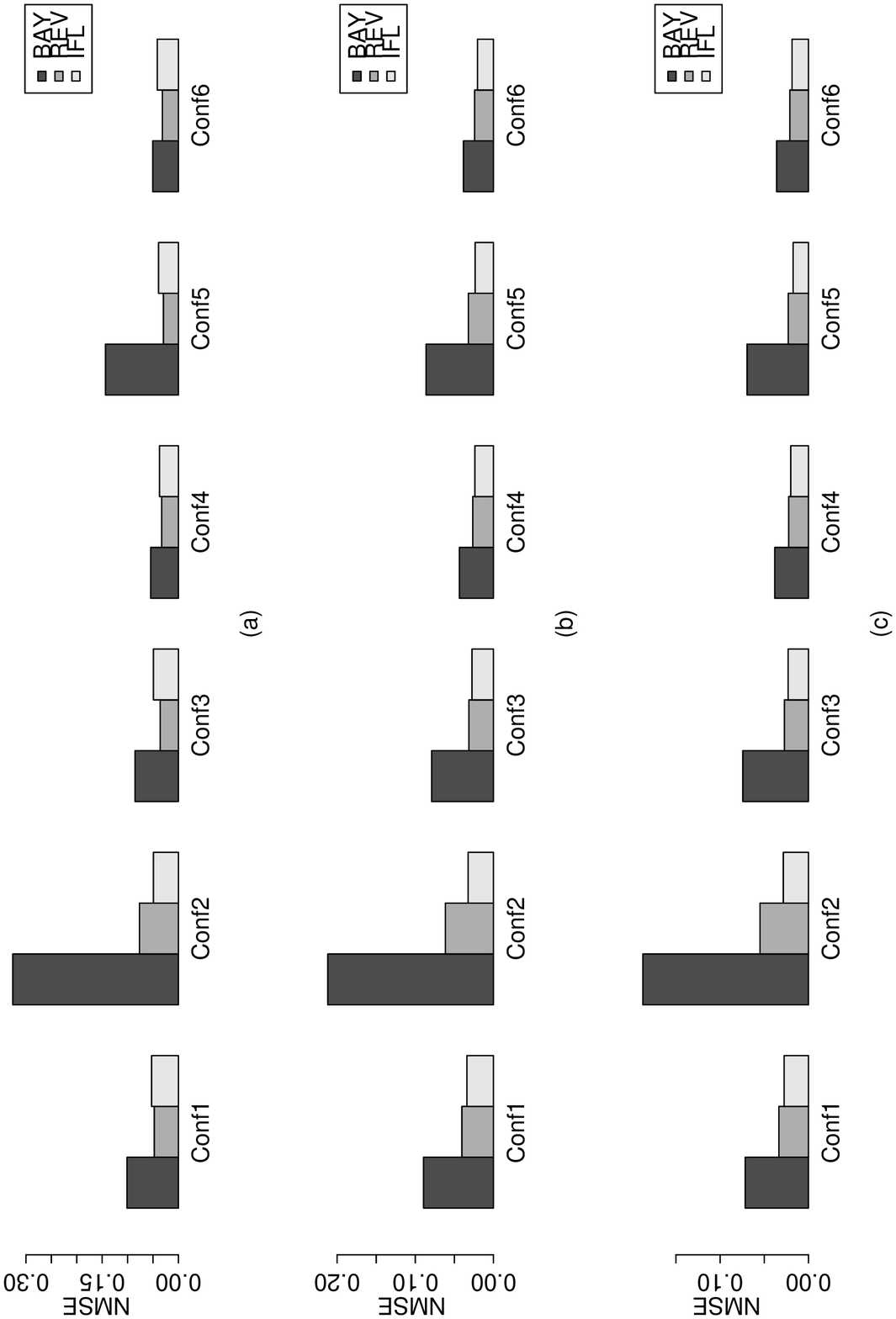}
  \caption{Evolution of the $NMSE$ for quantile $Q_{0.995}$ in
    function of the region configuration. Target site sample size: (a)
    10, (b) 25 and (c) 40.}
  \label{fig:globMSEQ995}
\end{figure*}

Figures~\ref{fig:globMSEQ75} to~\ref{fig:globMSEQ995} illustrate the
results for different target site sample sizes and regions. We
concentrate on the $NMSE$ criteria since it measures variation of the
estimator around the true parameter value.

From Figure~\ref{fig:globMSEQ75}, it is clear that Bayesian
estimations, i.e. $BAY$ and $REV$, of $Q_{0.75}$ are more accurate;
specially for a target site sample size of 10. For larger target site
sample sizes, Bayesian approaches are always more competitive than the
$IFL$ estimator, even if this is less clear-cut on the graphs.
Furthermore, $BAY$ and $REV$ estimators often have the same
performance. This result is logical as the $Q_{0.75}$ value is mostly
impacted by the location parameter $\mu$. Thus, reversible jumps do
not have a significant result on $REV$ $Q_{0.75}$ estimation.

The plots in Figure~\ref{fig:globMSEQ95} and those displayed in
Figure~\ref{fig:globMSEQ75} are quite different. For a target site
sample size of 10, both Bayesian approaches are the most accurate -
except for $BAY$ applied on $Conf2$ and $Conf5$ - and the $REV$
estimator leads always to the smallest $NMSE$. Thus, $REV$ is the most
competitive model. For larger target site sample sizes, $REV$ is at
least as accurate as $IFL$, except for $Conf2$.

For $Q_{0.995}$ and a target site sample size of 10, $REV$ is the most
accurate model, except for $Conf2$. As the target site sample size
increases, the $IFL$ approach becomes more efficient. However, for
these cases, $NMSE$ for the $REV$ estimator are often close to the
$IFL$ ones. Although the $BAY$ approach performs poorly for
$Q_{0.995}$, its $NMSE$ for $Conf6$ is close to the $REV$ and $IFL$
ones.

In conclusion, these results illustrate the good overall performance
of the $REV$ model. Indeed, this approach benefits from the efficiency
of the $BAY$ estimator for quantiles with small non exceedence
probabilities while being as competitive as the $IFL$ approach for
larger non exceedence probabilities.

However, the Bayesian approaches outperform the index flood model but
differences in accuracy seem to be less and less significant as the
sample site increases. This may be related to the white noise
introduce in the generation procedure. Indeed, this white noise is
independent of the target site sample size and may strongly penalize
the performances of the both Bayesian approaches. The next section
tries to outline the effect of the target site index flood estimation
error to the quantile estimates.

\section{Effect of Bias on the Target Site Index Flood Estimation}
\label{sec:effBias}

According to the model being considered, two types of biases are
encountered for the target site index flood estimation. Indeed, on one
hand, the index flood for the $IFL$ model is derived from the target
site sample. On the other hand, for $BAY$ and $REV$ approaches, the
index flood is estimated from a scaling model. Thus, biases on index
flood estimation are due to the relevance of this scaling model but
also to the index flood error estimation for the other sites within
the region.

\begin{figure*}
  \centering
  \includegraphics[angle=-90,width=\textwidth]{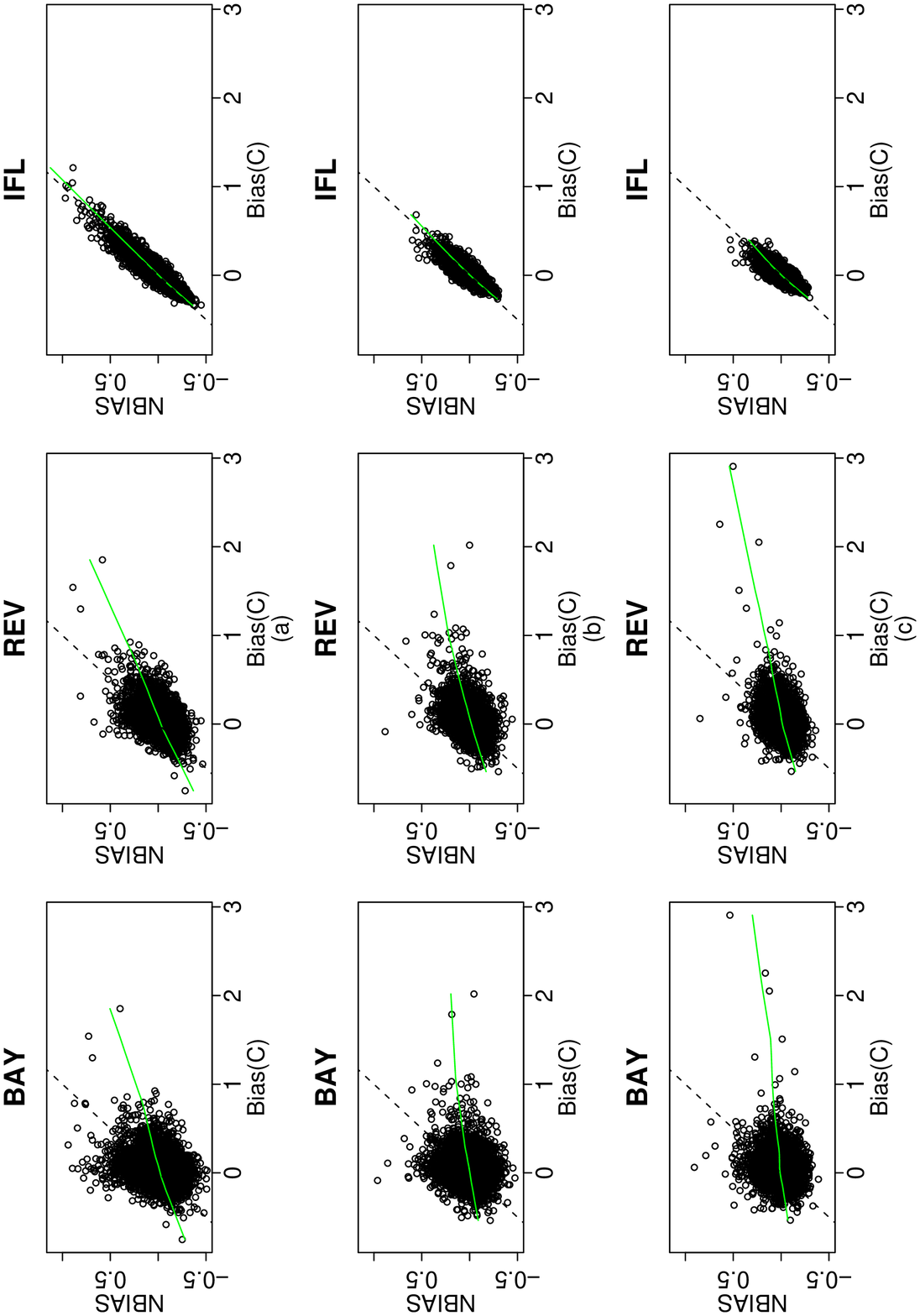}
  \caption{Evolution of $NBIAS$ for $Q_{0.95}$ in function of the
    normalized bias on target site index flood estimation (Bias(C)).
    Target site sample size: (a) 10, (b) 25 and (c) 40. Solid green
    lines: local smoothers, black dashed lines: $y = x$.}
  \label{fig:evNBIAS}
\end{figure*}

To illustrates these two types of biases, the normalized bias on
target site index flood estimation is computed as follows:

\begin{equation}
  \label{eq:bias(c)}
  Bias(C) = \frac{\hat{C} - C}{C}
\end{equation}
where $C$ is the target site index flood, and $\hat{C}$ is an estimate
of $C$. Figure~\ref{fig:evNBIAS} depicts changes in $NBIAS$ for
quantile $Q_{0.95}$ in function of $Bias(C)$. As normalized biases are
considered, statistics for the six configurations are plotted in the
same graphic. Solid lines correspond to local polynomial regression
fits to help underline trends.

Scatter-plots in Figure~\ref{fig:evNBIAS} show clearly these two types
of biases. Indeed, on one hand, the range of $Bias(C)$ is not the same
for $IFL$ than for $BAY$ and $REV$, particularly for a target site
sample size of 25 and 40. On the other hand, for the $BAY$ and $REV$
approaches, biases on index flood estimation are independent of the
target site sample size; while this is not the case for $IFL$. This
last point is also illustrated as the bias ranges for the Bayesian
approaches remain the same for all target site sample size. Thus, for
large sample size, efficiency of the Bayesian estimators may be too
much impacted as the artificial bias introduced in the generation
procedure is too penalizing.

The Bayesian approaches do not have the same behaviour than the $IFL$
model. In particular, $BAY$ and $REV$ seem to be less sensitive to a
large bias in target site index flood estimation. $NBIAS$ for the
$IFL$ model are clearly linear with a response $y=x$. This last point
is an expected result.  Indeed, apart from sampling variability, if a
unique regional distribution exists, quantile $IFL$ estimate biases
are only induced by biases on target site index flood estimates. Thus,
the relevance of the generation procedure is corroborated.

The main difference between the $BAY$ and $REV$ estimators is the
dispersion around local smoothers. Indeed, $REV$ has a smaller range
while preserving the same robustness to the bias on target site index
flood estimation.

These results and conclusions are independent of the target site index
flood estimation procedure. However, the performance of the two
Bayesian estimators is related to the bias and variance of the target
site index flood estimate. Thus, for similar variance, these results
should be identical if GAMs or Kriging were used.

\section{Suggestions for Region Configuration}
\label{sec:guidRegConf}

This section attempts to present some suggestions for building
suitable pooling groups according to the considered estimator.
\cite{Hosking1997} already advice not to build regions greater than 20
sites because of the small gain affected with additional stations.
However, they only focus on the $IFL$ methodology. We attempt to do
the same for the two Bayesian estimators considered in this study. For
this purpose,
tables~\ref{tab:guidRegConfIFL},~\ref{tab:guidRegConfBAY}
and~\ref{tab:guidRegConfREV} include the $NMSE$ and the related
standard errors for each configuration and target site sample size.


\begin{table*}
  \caption{Changes in $NMSE$ for $Q_{0.75}, Q_{0.95}$ and $Q_{0.995}$
    in function of the region configuration and the target site sample
    size for the $IFL$ estimator. Related standard errors are
    displayed in brackets.} 
  \label{tab:guidRegConfIFL}
  \centering
  \begin{tabular}{cccccccc}
    \hline
    \multirow{2}*{Model} & \multicolumn{3}{c}{Heavy Tail} & &
    \multicolumn{3}{c}{Light Tail}\\
    \cline{2-4} \cline{6-8}
    & $Conf1$ & $Conf2$ & $Conf3$ && $Conf4$ & $Conf5$ & $Conf6$\\
    \hline
    & \multicolumn{7}{c}{Target site sample size 10}\\
    $Q_{0.75}$ & 0.037 (3e-3)& 0.035 (3e-3)& 0.037 (3e-3)&& 0.025 
    (2e-3) & 0.029 (2e-3)& 0.029(3e-3)\\ 
    $Q_{0.95}$ & 0.038 (3e-3)&0.037 (3e-3)&0.039 (3e-3)&&0.027
    (4e-3)&0.031 (2e-3)&0.032 (3e-3)\\
    $Q_{0.995}$ & 0.053 (4e-3)&0.049 (3e-3)&0.049 (4e-3)&&0.037
    (2e-3)&0.039 (3e-3)&0.042 (4e-3)\\ 
    & \multicolumn{7}{c}{Target site sample size 25}\\
    $Q_{0.75}$ & 0.014 (8e-4)& 0.015 (1e-3)& 0.015 (1e-3)&& 0.011
    (7e-4)& 0.011 (7e-4)& 0.011(7e-4)\\ 
    $Q_{0.95}$ & 0.018 (1e-3)&0.018 (1e-3)&0.018 (1e-3)&&0.014
    (9e-4)&0.014 (9e-4)&0.013 (9e-4) \\
    $Q_{0.995}$ & 0.034 (2e-3)&0.032 (2e-3)&0.027 (2e-3)&&0.024
    (2e-3)&0.023 (2e-3)&0.020 (1e-3)\\ 
    & \multicolumn{7}{c}{Target site sample size 40}\\
    $Q_{0.75}$ & 0.010 (6e-4)&0.009 (6e-4)&0.010 (6e-4)&&0.007
    (4e-4)&0.007 (4e-4)&0.007 (5e-4)\\
    $Q_{0.95}$ & 0.013 (8e-4)&0.013 (8e-4)&0.012 (8e-4)&&0.010
    (6e-4)&0.009 (5e-4)&0.010 (6e-4)\\ 
    $Q_{0.995}$ & 0.028 (2e-3)&0.028 (2e-3)&0.023 (2e-3)&&0.020
    (1e-3)&0.017 (1e-3)&0.019 (1e-3)\\ 
    \hline
  \end{tabular}
\end{table*}  

From Table~\ref{tab:guidRegConfIFL}, the $IFL$ estimator seems to have
the same performance level independently of the configuration. This
result points out that the information is not used optimally as
regions with the most information (i.e. $Conf3$ and $Conf6$) do not
always lead to better estimations. This last point corroborates a
previous comments of~\cite{Ribatet2007a}.


\begin{table*}
  \caption{Changes in $NMSE$ for $Q_{0.75}, Q_{0.95}$ and $Q_{0.995}$
    in function of the region configuration and the target site sample
    size for the $BAY$ estimator. Related standard errors are
    displayed in brackets.}
  \label{tab:guidRegConfBAY}
  \centering
  \begin{tabular}{cccccccc}
    \hline
    \multirow{2}*{Model} & \multicolumn{3}{c}{Heavy Tail} & &
    \multicolumn{3}{c}{Light Tail}\\
    \cline{2-4} \cline{6-8}
    & $Conf1$ & $Conf2$ & $Conf3$ && $Conf4$ & $Conf5$ & $Conf6$\\
    \hline
    & \multicolumn{7}{c}{Target site sample size 10}\\
    $Q_{0.75}$ & 0.015 (1e-3)& 0.015 (1e-3)& 0.012 (8e-4)&& 0.011
    (6e-4)& 0.012 (9e-4)& 0.011(9e-4)\\ 
    $Q_{0.95}$ & 0.035 (2e-3)&0.063 (4e-3)&0.030 (2e-4)&&0.022
    (1e-3)&0.037 (3e-3)&0.023 (2e-3)\\
    $Q_{0.995}$ & 0.101 (1e-2)&0.326 (3e-2)&0.085 (6e-3)&&0.054
    (5e-3)&0.144 (1e-2)&0.050 (3e-3)\\ 
    & \multicolumn{7}{c}{Target site sample size 25}\\
    $Q_{0.75}$ & 0.010 (6e-4)& 0.011 (7e-4)& 0.009 (5e-4)&& 0.008
    (5e-4)& 0.007 (4e-4)& 0.007(5e-4)\\ 
    $Q_{0.95}$ & 0.026 (2e-3)&0.041 (3e-3)&0.025 (1e-3)&&0.017
    (1e-3)&0.023 (1e-3)&0.016 (9e-4) \\ 
    $Q_{0.995}$ & 0.089 (8e-3)&0.212 (2e-2)&0.079 (4e-3)&&0.044
    (3e-3)&0.086 (6e-3)&0.038 (2e-3)\\
    & \multicolumn{7}{c}{Target site sample size 40}\\
    $Q_{0.75}$ & 0.008 (5e-4)&0.008 (5e-4)&0.007 (4e-4)&&0.005
    (3e-4)&0.005 (3e-4)&0.006 (4e-4)\\
    $Q_{0.95}$ & 0.020 (1e-3)&0.032 (2e-3)&0.020 (1e-3)&&0.012
    (8e-4)&0.015 (9e-4)&0.013 (8e-4)\\
    $Q_{0.995}$ & 0.072 (5e-3)&0.187 (2e-2)&0.074 (5e-3)&&0.038
    (3e-3)&0.070 (6e-3)&0.036 (2e-3)\\ 
    \hline
  \end{tabular}
\end{table*}

Table~\ref{tab:guidRegConfBAY} shows that the $BAY$ estimator is more
accurate with ``medium'' regions, i.e. $Conf3$ and $Conf6$. However,
results for ``small'' regions, i.e. $Conf1$ and $Conf4$, are often
close to the best ones - especially for a light tail. Thus, it is
preferable to work with well-instrumented sites, i.e. $Conf1, Conf3,
Conf4$ and $Conf6$.


\begin{table*}
  \caption{Changes in $NMSE$ for $Q_{0.75}, Q_{0.95}$ and $Q_{0.995}$
    in function of the region configuration and the target site sample
    size for the $REV$ estimator. Related standard errors are
    displayed in brackets.} 
  \label{tab:guidRegConfREV}
  \centering
  \begin{tabular}{cccccccc}
    \hline
    \multirow{2}*{Model} & \multicolumn{3}{c}{Heavy Tail} & &
    \multicolumn{3}{c}{Light Tail}\\
    \cline{2-4} \cline{6-8}
    & $Conf1$ & $Conf2$ & $Conf3$ && $Conf4$ & $Conf5$ & $Conf6$\\
    \hline
    & \multicolumn{7}{c}{Target site sample size 10}\\
    $Q_{0.75}$ & 0.014 (1e-3)& 0.011 (7e-4)& 0.011 (7e-4)&& 0.011
    (6e-4) & 0.010 (7e-4)& 0.011(9e-4)\\ 
    $Q_{0.95}$ & 0.026 (2e-3)&0.024 (2e-3)&0.019 (1e-3)&&0.018
    (1e-3)&0.016 (1e-3)&0.019 (2e-3)\\
    $Q_{0.995}$ & 0.047 (3e-3)&0.077 (2e-2)&0.036 (2e-3)&&0.033
    (2e-3)&0.030 (2e-3)&0.032 (3e-3)\\ 
    & \multicolumn{7}{c}{Target site sample size 25}\\
    $Q_{0.75}$ & 0.009 (6e-4)& 0.009 (6e-4)& 0.008 (5e-4)&& 0.008
    (5e-4)& 0.006 (4e-4)& 0.007(5e-4)\\ 
    $Q_{0.95}$ & 0.019 (1e-3)&0.020 (2e-3)&0.016 (9e-4)&&0.014
    (1e-3)&0.014 (9e-4)&0.014 (9e-4) \\ 
    $Q_{0.995}$ & 0.040 (3e-3)&0.061 (1e-2)&0.031 (2e-3)&&0.026
    (2e-3)&0.032 (3e-3)&0.024 (2e-3)\\ 
    & \multicolumn{7}{c}{Target site sample size 40}\\
    $Q_{0.75}$ & 0.008 (5e-4)&0.007 (5e-4)&0.006 (4e-4)&&0.005
    (3e-4)&0.005 (3e-4)&0.006 (3e-4)\\
    $Q_{0.95}$ & 0.015 (1e-3)&0.016 (1e-3)&0.012 (9e-4)&&0.010
    (7e-4)&0.010 (5e-4)&0.011 (6e-4)\\ 
    $Q_{0.995}$ & 0.034 (2e-3)&0.055 (1e-2)&0.027 (2e-3)&&0.022
    (2e-3)&0.023 (2e-3)&0.021 (1e-3)\\ 
    \hline
  \end{tabular}
\end{table*}  

Table~\ref{tab:guidRegConfREV} shows that the $REV$ estimator more
efficient with ``medium'' regions, i.e. $Conf3$ and $Conf6$. In
addition, it seems to be more accurate with few but well-instrumented
gauging stations rather more but less-instrumented ones. Nevertheless
for a light tail, all configurations seems to lead to similar
performance levels.

Tables~\ref{tab:guidRegConfIFL},~\ref{tab:guidRegConfBAY}
and~\ref{tab:guidRegConfREV} show that the estimation of $Q_{0.75}$ is
independent of the region configuration for all estimators. Thus, it
seems that the regional information is not relevant for quantiles with
small non exceedence probabilities.

\section{Conclusions}
\label{sec:disc}

This article introduced a new Bayesian estimator which uses regional
information in an innovative way. The proposed model accounts for a
fixed regional shape parameter with a non null probability. Thus, as
in~\cite{Ribatet2007a}, the regional information is still used to
elicit the prior distribution. However, the prior distribution is now
a mixture of a GEV/GPD and a GEV/GPD with only two parameters - the
remaining one corresponds to the fixed regional shape parameter.

The estimation procedure is achieved using reversible jump Markov
chains~\citep{Green1995}; and theoretical details for simulated suited
Markov chains were presented. A sensitivity analysis for the proposed
algorithm was performed. The results showed that the estimates are
consistent provided that the probability attributed to the fixed
regional shape parameter is positive. In addition, as noticed
by~\cite{Stephenson2004a}, the credibility intervals are sensitive to
this probability value. Thus, the proposed estimator relates this
probability value to the homogeneity degree of the region - using the
heterogeneity statistic of~\cite{Hosking1997}. Therefore, the
credibility intervals take into account the belief about the fixed
regional shape parameter to be the true value.

A performance analysis was carried out on stochastic data for three
different estimators. For this purpose, another algorithm which
generates stochastic homogeneous regions was implemented. The good
overall performance of the proposed estimator has been demonstrated.
Indeed, on one hand, this approach combines the accuracy of the
regional Bayesian approach of~\cite{Ribatet2007a} for quantiles
associated to small exceedence probabilities. On the other hand, the
duality of the prior distribution (and the fixed regional shape
parameter) allows the proposed estimator to be at least as efficient
as the index flood model. Thus, this new estimator seems very suited
for regional estimation when the target site is not well instrumented.

Furthermore, the two Bayesian approaches considered here appear to be
less sensitive to biases on target site index flood estimation than
the index flood estimator. Thus, the Bayesian approaches are more
readily adaptable which is a major advantage as errors on the index
flood estimation are often uncontrollable.

As noticed by~\cite{Ribatet2007a}, the index flood model does not use
information optimally. This point is corroborated in this study as the
model initiated by~\cite{Dalrymple1960} is not inevitably more
accurate as the information within the pooling group increases. This
is not the case for the Bayesian approaches. In addition, they seem to
be more accurate when dealing with regions with well instrumented
sites, particularly for large quantiles.

All statistical analysis were carried out by use of~\cite{Rsoft}. For
this purpose, the algorithm presented in section~\ref{subsec:postEst}
was incorporated in the \textbf{evdbayes}
packages~\citep{Stephenson2006}. The algorithm for the generation
procedure is available on request from the author.

\begin{acknowledgments}
  The authors wish to thank Alec Stephenson for providing the original
  codes of his article. The financial support provided by the National
  Science and Engineering Research Council of Canada (NSERC) is
  acknowledge. We are also grateful to the editor, the associate
  editor and two anonymous referees for useful criticism of the
  original version of the paper.
\end{acknowledgments}

\appendix
\section{The Metropolis-Hastings Algorithm}
\label{sec:metro}

In this section, the Metropolis-Hastings algorithm is presented.
According to the results derived by \citet{Green1995}, some details
will be given to consider the reversible jump case. The basic idea of
the Metropolis-Hastings algorithm is to obtain a Markov chain that
converges to a known stationary distribution. The strength of the
Metropolis-Hasting approach is that the convergence is reached
whatever the initial state of the Markov chain is and that the
distributions could be known up to a constant.

Let $f$ denote the target distribution of interest. Most often, in
Bayesian inference, $\pi$ will be the posterior distribution for the
parameters. Let $q(\cdot, x)$ be the proposal distribution i.e. the
proposal states will be sampled from this proposal distribution given
the current state $x_t$. The Metropolis-Hastings algorithm can be
summarized as follows:

\begin{enumerate}
\item Generate $u$ from a uniform distribution on $[0,1]$;
\item Generate $x_\mathrm{prop}$ from $q(\cdot, x_t)$
\item $\Delta_\mathrm{class} \leftarrow
  \frac{f(x_\mathrm{prop})}{f(x_t)} \frac{q(x_t
    | x_\mathrm{prop})}{q(x_\mathrm{prop} | x_t)}$
\item \textbf{if} $u < \min\left(1, \Delta_\mathrm{class}\right)$
  \textbf{then}
\item $\quad x_{t+1} \leftarrow x_\mathrm{prop}$
\item \textbf{else}
\item $\quad x_{t+1} \leftarrow x_t$
\item \textbf{endif}
\item Go to 1.
\end{enumerate}

The initial Metropolis-Hastings algorithm can not account for
dimensional switch. For this purpose, the ``jumps'' between sub-spaces
must be defined (see
equations~\eqref{eq:proploc1}--\eqref{eq:propshape1} and
\eqref{eq:proploc2}--\eqref{eq:propshape2}) and the quantity
$\Delta_\mathrm{class}$ must be redefined each time a jump is
considered. Here, only a simple case of the reversible jumps approach
is considered (see Section 3.3 of \citet{Green1995}). If only two
moves $m_1(x_t)$ and $m_2(x_t)$ can occur with probabilities $p_1$ and
$p_2$ respectively, then the quantity $\Delta_\mathrm{class}$ must be
replaced by $\Delta_\mathrm{rev}$. Consequently, for a proposal move
of type $m_1$ :

\begin{equation}
  \label{eq:deltaREV}
  \Delta_\mathrm{rev} = \Delta_\mathrm{class} \frac{p_1}{p_2}  J_1
\end{equation}
where $J_1$ is the jacobian of the transformation $x_t \mapsto
m_1(x_t)$. If the proposal move is of type $m_2$, then

\begin{equation}
  \label{eq:deltaREV2}
  \Delta_\mathrm{rev} = \Delta_\mathrm{class} \frac{p_2}{p_1}  J_2
\end{equation}
where $J_2$ is the jacobian of the transformation $x_t \mapsto
m_2(x_t)$.

\end{article}



\end{document}